\documentclass[10pt, oneside]{article}   	% use "amsart" instead of "article" for AMSLaTeX format
\usepackage{geometry}                		% See geometry.pdf to learn the layout options. There are lots.
\usepackage{graphicx}				% Use pdf, png, jpg, or eps? with pdflatex; use eps in DVI mode
\usepackage{amssymb}
\usepackage{cite}
\usepackage{color}
\usepackage{amsmath}
\usepackage{hyperref}

\usepackage{subcaption}
\usepackage{graphicx}
\usepackage[all]{xy}
\usepackage{float}
\usepackage{parskip}
\usepackage{tikz}

\usetikzlibrary{arrows,positioning,automata,shadows,fit,shapes,calc,math,decorations.markings,decorations.pathreplacing}

\title{Overcoming timestep limitations in boosted-frame Particle-In-Cell simulations of plasma-based acceleration}

\author{
Olga Shapoval\textsuperscript{a}\thanks{CONTACT O.~Shapoval. Email: oshapoval@lbl.gov},
Remi Lehe\textsuperscript{a}\thanks{CONTACT R.~Lehe. Email: rlehe@lbl.gov},
Maxence Th\'evenet\textsuperscript{a}\thanks{Now at DESY, 22607 Hamburg, Germany},\\
Edoardo Zoni\textsuperscript{a},
Yinjian Zhao\textsuperscript{a},
Jean-Luc Vay\textsuperscript{a}
\\
\textsuperscript{a}Lawrence Berkeley National Laboratory, Berkeley, CA 94720, USA
}

\begin{document}

\tikzset{
->-/.style={decoration={markings,mark=at position #1 with {\arrow{>}}},postaction={decorate}}
}

\maketitle

%\begin{keywords}

\begin{abstract}
Explicit electromagnetic Particle-In-Cell (PIC) codes are typically limited by the
Courant-Friedrichs-Lewy (CFL) condition, which implies that the timestep
multiplied by the speed of light must be smaller than the smallest cell size.
In the case of boosted-frame PIC
simulations of plasma-based acceleration, this limitation can be a major hinderance
as the cells are often very elongated along the longitudinal direction and the
timestep is thus limited by the small, transverse cell size. This entails many small-timestep PIC iterations, and can limit the potential speed-up
of the boosted-frame technique.
Here, by using a CFL-free analytical spectral solver, and by mitigating
additional numerical instabilities that arise at large timestep, we show that
it is possible to overcome traditional limitations on the timestep and thereby
realize the full potential of the boosted-frame technique over a much wider range of parameters.
\end{abstract}

\renewcommand{\vec}{\boldsymbol}

\newcommand{\kj}{\vec{k}_{\vec{j}}}
\newcommand{\Vol}{\mathcal{V}}
\newcommand{\xm}{\vec{x}_{\vec{m}}}
\newcommand{\Knu}{\vec{K}_{\vec{\nu}}}
\newcommand{\Xnu}{\vec{X}_{\vec{\nu}}}
\newcommand{\spectral}[1]{\hat{\mathcal{#1}}}
\newcommand{\Spectral}[1]{\vec{\hat{\mathcal{#1}}}}
\newcommand{\kx}{\hat{k}_x}
\newcommand{\kz}{\hat{k}_z}
\newcommand{\K}{\vec{\hat{k}}}
\newcommand{\Ex}{\spectral{E}_x}
\newcommand{\Ez}{\spectral{E}_z}
\newcommand{\By}{\spectral{B}_y}
\newcommand{\Fx}{\spectral{F}_x}
\newcommand{\Fz}{\spectral{F}_z}
\newcommand{\cotan}{\mathrm{cot}}
\newcommand{\watch}[1]{\textcolor{red}{#1}}

%\section*{Main}

Particle-In-Cell (PIC) simulations \cite{Hockney1988,Birdsall2004} are key to the development of plasma-based accelerators and of their potential future applications \cite{AlbertNJP2020}.
However, these simulations can typically be very computationally expensive. %, especially in the case of laser-wakefield acceleration where the small laser wavelength imposes strong demands on the numerical resolution.
One way to reduce their computational cost is to use the boosted-frame technique \cite{VayPRL2007}, whereby the simulation is performed in a Lorentz frame moving relativistically in the same direction as the beam or laser driver. The boosted-frame technique is nowadays routinely used in simulations of plasma-based accelerators, and can speed up simulations by several orders of magnitude. To a large extent, this was made possible by the development of a number of algorithms that mitigate the Numerical Cherenkov Instability (NCI) \cite{GodfreyJCP1974,GodfreyJCP1975,VayJCP2011,GodfreyJCP2013,XuCPC2013,GodfreyJCP2014,GodfreyJCP2014b,GodfreyIEEE2014,GodfreyCPC2015,YuCPC2015,YuCPC2015-Circ,LehePRE2016,KirchenPoP2016,KirchenPRE2020,FeiCPC2017,PukhovJCP2020} -- a numerical instability that would otherwise rapidly grow in the boosted frame and irremediably corrupt the simulated physics.

The remarkable speedup afforded by the boosted-frame technique is due largely to the possibility of increasing the timestep in the boosted frame, and thereby reducing the number of PIC iterations to be performed compared to a corresponding laboratory-frame simulation.
For example, in a typical laboratory-frame simulation of laser-wakefield acceleration, both the longitudinal cell size $\Delta z_{lab}$ and timestep $\Delta t_{lab}$ are constrained to resolve the small laser oscillations at wavelength $\lambda_{lab}$: $\Delta z_{lab} \ll \lambda_{lab}$ for a laser propagating along $z$, with $c\Delta t_{lab}\leq \Delta z_{lab} \ll \lambda_{lab}$ (while the transverse cell size is usually much larger: $\Delta x, \Delta y \gg \Delta z_{lab}$). By contrast, in a Lorentz boosted frame drifting along $z$ at relativistic velocity with a Lorentz factor $\gamma_{b} \gg 1$, the laser oscillations are dilated by a factor of approximately $2\gamma_{b}$ ($\lambda \approx 2\gamma_{b} \lambda_{lab}$), which greatly relaxes the constraints on the longitudinal cell size and timestep: $\Delta z \ll 2\gamma_{b} \lambda_{lab}$, with $c\Delta t \leq \Delta z \ll 2\gamma_{b} \lambda_{lab}$ (where the quantities $\Delta z$ and $\Delta t$ denote the longitudinal cell size and timestep \emph{in the boosted frame}).

However, for large $\gamma_{b}$, as the constraints imposed by the laser are relaxed, the timestep often becomes constrained instead by the transverse cell size: $c\Delta t \leq \Delta x, \Delta y$. (Note that the transverse cell size is left unchanged  in the boosted-frame simulation as compared to the corresponding laboratory-frame simulation, since transverse physical length scales are unchanged by the Lorentz transform.)
In the case of Finite-Difference Time-Domain (FDTD) PIC algorithms, this constraint on the timestep is due to the Courant-Friedrichs-Lewy (CFL) condition \cite{CFL,Taflove}. Similarly, the Pseudo-Spectral Time-Domain (PSTD) PIC algorithm \cite{LiuMOTL1997} also has a CFL condition. As a consequence of the CFL condition, the timestep of the boosted-frame simulation is relatively small and limits the potential computational speedup, even though the physics at stake does not necessarily require such a high temporal resolution.

On the other hand, unlike FDTD and PSTD PIC algorithms, Pseudo-Spectral Analytical Time-Domain (PSATD) PIC algorithms \cite{Haber,VayJCP2013},
which integrate analytically Maxwell's equations over one time step
in Fourier space, do not have a similar CFL condition.
It follows that boosted-frame PIC simulations that use the PSATD Maxwell
solver could use a larger timestep, as it is thus not explicitly constrained by the transverse resolution.
However, it turns out that PSATD boosted-frame simulations are empirically unstable for $c\Delta t > \Delta x, \Delta y$.
More specifically, the Galilean PSATD algorithm \cite{LehePRE2016,KirchenPoP2016,KirchenPRE2020}, which does efficiently mitigate the NCI for $c\Delta t < \Delta x, \Delta y$, does not seem to suppress the NCI anymore for $c\Delta t > \Delta x, \Delta y$.

This paper examines the nature of this resurgent NCI and shows that this instability can be strongly mitigated with a new algorithm, referred to as the \emph{averaged Galilean PSATD}, whereby a key feature of the PSATD algorithm is exploited to analytically average the electromagnetic fields in time before gathering them onto the macroparticles. Hence, with this new algorithm, simulations can run with a large timestep ($c\Delta t  \leq\Delta z \gg \Delta x, \Delta y$) and exhibit the corresponding computational speedup, while preserving the integrity of the simulated physics.
While this development was motivated here with the example of laser-wakefield acceleration, it is generally applicable to any simulation where the physics imposes a high transverse spatial resolution but does not impose such strong constraints on the timestep, so that it would be advantageous to use a large timestep compared to the cell size. For instance, this also includes the simulations of low-emittance pencil-like beams \cite{MehrlingIEEE2018}, in which the space charge requires a high transverse resolution, but has a relatively slow time evolution.

The remainder of the paper is structured as follows. We first examine in more
detail the NCI that occurs for large timesteps in the case of the standard
Galilean PSATD algorithm. Based on this analysis, we introduce the averaged
Galilean PSATD algorithm and describe the corresponding modified PIC loop.
We then demonstrate the stability of this new algorithm with large timesteps,
first for a uniform plasma, and then for 2D simulations of laser-wakefield
acceleration (LWFA) and 3D simulations of plasma wakefield acceleration (PWFA).

\section*{Results}
\subsection*{Limitations of the standard Galilean PSATD algorithm for large timesteps}
\label{sec:standardGalilean}
As mentioned in the introduction, boosted-frame simulations with the Galilean PSATD algorithm are typically unstable when using a large timestep $c\Delta t > \Delta x, \Delta y$.
Here, we illustrate this by examining the theoretical NCI growth rate of the Galilean PSATD algorithm for two-dimensional simulations of a uniform plasma drifting at a relativistic velocity $\vec{v_0} =v_0\vec{u}_z$ (where $\vec{u}_z$ is the unit vector along the $z$ axis). As a reminder, the Galilean PSATD algorithm solves the Maxwell equations on a moving grid, which drifts at a velocity $\vec{v}_{gal} = v_{gal} \vec{u}_z$. This algorithm was shown to suppress the NCI when $v_{gal} = v_0$ \cite{LehePRE2016}.

In this section, we in fact consider two cases: that of a matching velocity $v_{gal} = v_0$ and that of a slightly detuned velocity $v_{gal} = 0.99 \,v_0$. Conceptually, these two cases represent -- at a simplified level -- different areas of the simulation box, in the case of a realistic LWFA simulation. More specifically, the matched case ($v_{gal} = v_0$) represents the background, quiescent plasma, far from the driver and the wakefield, since the Galilean velocity is typically chosen to match its velocity (\textsl{i.e.}, $v_{gal} = -\sqrt{1 - 1/\gamma_{b}^2} \,c$). On the other hand, the case of the detuned velocity represents the perturbed plasma around the laser driver and in the wakefield, where the local velocity is different than that of the background plasma, and thus different than the Galilean velocity.

In both of these cases, we choose $\Delta z \gg \Delta x$. This is typical for boosted-frame simulations with a large $\gamma_{b}$, since the longitudinally Lorentz-dilated driver and wakefield relax the requirement on the longitudinal resolution. We then further consider two cases: that of a small timestep $c\Delta t = \Delta x$ and that of a large timestep $c\Delta t = \Delta z$. Note that the latter case would not be allowed by the CFL condition of an FDTD algorithm.

\begin{figure}[htb!]
\centering
\begin{subfigure}{0.95\linewidth}
\includegraphics[width=\linewidth]{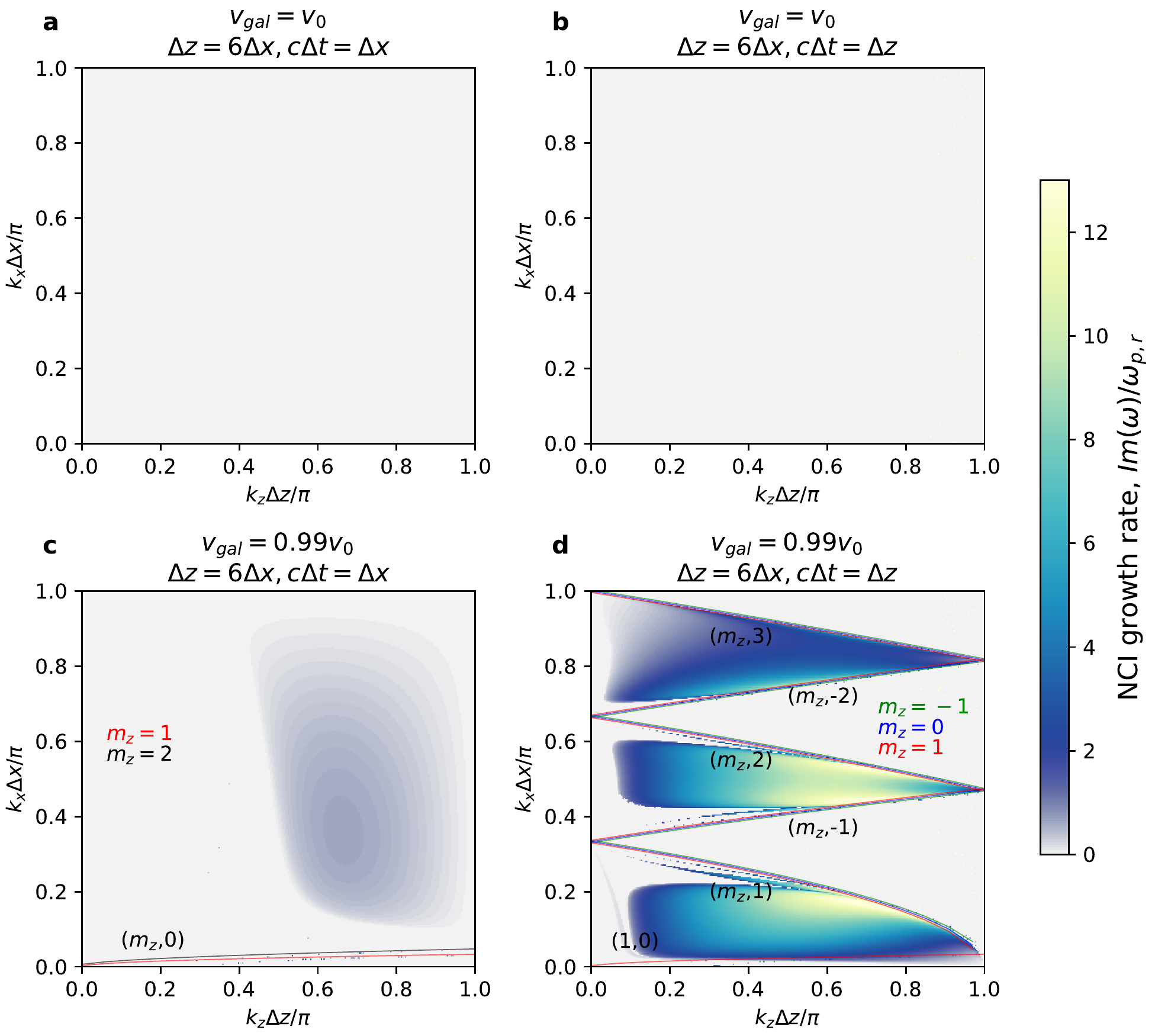}
\end{subfigure}
\caption{\textbf{NCI growth rate of the Galilean PSATD scheme.} Normalized NCI growth rate  $Im(\omega)/\omega_{p,r}$ in spectral  ($k_x$,$k_z$) space, calculated from the analytical stability analysis for different Galilean velocities $v_{gal} = v_0$ (a)-(b) and $v_{gal} = 0.99 \,  v_0$ (c)-(d), and for different timesteps $c\Delta t =\Delta x$ (left) and $c\Delta t =6 \Delta x$ (right). The solid colored lines correspond to well-known aliased NCI resonant modes, with alias number $(m_z,n)$, as given by equation~(\ref{eq:res_modes}). In this simulation, a uniform plasma drifts at a velocity $v_0 = c ( 1 - 1/\gamma_b^2 )^{1/2}$  with $\gamma_b = 130$, and the transverse and longitudinal cell sizes are $\Delta x =6.4 \times 10^{-2} \, k_{p,r}^{-1}$ and $\Delta z = 6 \Delta x$, respectively (where $k_{p,r}^2 = n_0e^2/(m_e\epsilon_0c^2 \gamma_0)$, and where $n_0$ is the plasma density).}
\label{fig:nci_growth_rate_gal}
\end{figure}

Fig.~\ref{fig:nci_growth_rate_gal} displays the theoretical NCI growth rate for the four possible combinations (\textsl{i.e.}, small/large timestep and matched/detuned Galilean velocity). The growth rates are obtained by solving the theoretical dispersion relation, namely equation~(19) in \cite{LehePRE2016}.
In order to guide the interpretation of this figure, we also plot the position of well-known NCI resonant
modes \cite{GodfreyJCP2013}, which are caused by temporal and spatial aliasing. For the Galilean PSATD algorithm, the equation of these aliased resonant modes
is given by:
\begin{equation}
\centering
k_{x,res} = \sqrt{ \Big(k_z\frac{v_0}{c} + m_z\frac{2\pi}{\Delta z}\frac{(v_0 - v_{gal})}{c} - \frac{2\pi n}{c\Delta t} \Big)^2 - k_z^2} \,, \quad  m_z, n\in \mathbb{Z}
\label{eq:res_modes}
\end{equation}
where $m_z$ is the spatial alias index and $n$ is the temporal alias index \cite{KirchenPRE2020}. As one can observe, if $v_{gal} \approx v_0$, the term proportional to $m_z$ almost cancels and the position of these lines mainly depends on the time aliasing $n$.

As can be seen in Fig.~\ref{fig:nci_growth_rate_gal} (a)-(b), in the matched-velocity case ($v_{gal}=v_0$), the Galilean PSATD algorithm suppresses the NCI, both for $c\Delta t = \Delta x$ (upper left panel) and $c\Delta t \gg \Delta x$ (upper right panel). By contrast, in the detuned case ($v_{gal} \neq v_0$, lower panels in Fig.~\ref{fig:nci_growth_rate_gal}), the NCI has a more noticeable growth rate. This growth rate is relatively small for $c\Delta t = \Delta x$, but is much larger for $c\Delta t \gg \Delta x$. In practice, this implies that the Galilean algorithm is relatively robust to velocity perturbations (e.g. in the wakefield) in the case of a small timestep ($c\Delta t \le \Delta x$), but it is much less robust to those perturbations in the case of a large timestep ($c\Delta t \gg \Delta x$).
This explains the empirical observation, mentioned in the introduction, that boosted-frame simulations of LWFA are typically unstable with large timestep.

Furthermore, in the case $v_{gal} \neq v_0$ and $c\Delta t \gg \Delta x$ (panels (c)-(d) in Fig.~\ref{fig:nci_growth_rate_gal}), we see that the large NCI growth rate is concentrated near time-aliased resonances. Thus, the NCI arises here primarily from the resonant interaction of particles with electromagnetic modes that are not resolved in time.
More specifically, the electromagnetic modes are in principle oscillating in time (as predicted by the analytical formulas used in the derivation of the PSATD algorithm), and in most cases their net effect on particles averages to zero. Yet because these fields are sampled with a (large) discrete timestep -- which can be comparable to the period of their oscillations -- particles may in certain cases see an almost constant value (because of aliasing) instead of an oscillating one, and thus experience a lasting, resonant effect.

\subsection*{Averaged Galilean PSATD algorithm}
\label{sec:averagedGalilean}

The above analysis suggests a natural remedy: when pushing the particles, instead of using the value of the fields sampled at a specific time $t = n\Delta t$ for some integer $n$, the particles should instead be pushed with the fields \emph{averaged in time} between $t = (n-1/2) \Delta t$ and $t = (n+1/2)\Delta t$. Averaging the fields over one timestep will barely affect the physics, provided that it is well-resolved in time with the chosen timestep. On the other hand, this average will damp the under-resolved modes that are spuriously resonant in Fig.~\ref{fig:nci_growth_rate_gal}.
Since this is a \emph{temporal} average, not a \emph{spatial} one, it will not affect the above-mentioned fine spatial details that typically impose a high transverse resolution $\Delta x$, $\Delta y$ (for example, the space-charge field of a low-emittance beam), as long as they vary slowly in time (in comparison to the time step used in the simulation).

\begin{figure}[htb!]
\centering
\begin{tikzpicture}
\def \Dt{3.2}
\def \yspac{1.}
% Axes
\draw[very thick,<->,red,>=stealth] (3.1*\Dt,0) -- (4.9*\Dt,0);
\draw (4*\Dt,0) node[red,anchor=south east]{Averaging interval};
\draw[->,>=stealth] (2*\Dt,0) -- (5.5*\Dt,0) node[anchor=north east]{Time};
\draw[dashed] (1.5*\Dt,0) -- (2*\Dt,0);
% Labels
\draw (1.6*\Dt,3.6*\yspac) node{\textbf{Spectral grid ($\vec{k}$)}};
\draw (1.6*\Dt,\yspac) node{\textbf{Spatial grid ($\vec{x}$)}};
\draw (1.6*\Dt,-\yspac) node{\textbf{Macroparticles}};
% Timesteps
\foreach \n in {3,5}
\draw[fill=white] (\n*\Dt,0) circle(0.1);
\draw[fill=black] (2*\Dt,0) circle(0.1) node[gray,anchor=north east]{$(n-1)\Delta t$};
\draw[fill=black] (4*\Dt,0) circle(0.1) node[gray,anchor=north east]{$n\Delta t$};;
% Fields
% -- n-1
\draw (1.8*\Dt,3*\yspac) node[black,fill=white,draw=black,rounded corners]{$\spectral{\rho}^{n-1}$};
\draw (2.25*\Dt,3*\yspac) node[black,fill=white,draw=black,rounded corners]{$\spectral{E}^{n-1}, \spectral{B}^{n-1}$};
% -- n-1/2
\draw (3*\Dt,3*\yspac) node[black,fill=white,draw=black,rounded corners]{$ \Spectral{J}^{n-1/2} $};
\draw (3*\Dt,-\yspac) node[black,fill=white,draw=black,rounded corners]{$\vec{p}^{n-1/2}$};
% -- n
\draw (3.6*\Dt,3*\yspac) node[black,fill=white,draw=black,rounded corners]{$\spectral{\rho}^{n}$};
\draw (3.9*\Dt,3*\yspac) node[blue,fill=white,draw=blue,rounded corners]{$  \Spectral{E}^{n}, \Spectral{B}^{n}$};
\draw (4.4*\Dt,3*\yspac) node[blue,fill=white,draw=blue,rounded corners]{$  \langle\Spectral{E}\rangle^{n}, \langle\Spectral{B}\rangle^{n}$};
\draw (4*\Dt,\yspac) node[blue,fill=white,draw=blue,rounded corners]{$\langle \vec{E}\rangle^{n}, \langle \vec{B}\rangle^{n}$};
\draw (4*\Dt,-\yspac) node[black,fill=white,draw=black,rounded corners]{$\vec{x}^{n}$};
\draw (4*\Dt,-1.6*\yspac) node[blue!50!red,fill=white,draw=blue!50!red,rounded corners]{$\langle\vec{E}\rangle(\vec{x}^{n}), \langle\vec{B}\rangle(\vec{x}^n)$};
% -- n+1/2
\draw (5*\Dt,-\yspac) node[blue!50!red,fill=white,draw=blue!50!red,rounded corners]{$\vec{p}^{n+1/2}$};
% a) Particle push
\draw[->,blue!50!red,>=stealth,thick] (3*\Dt,-1.4*\yspac) .. controls (3.2*\Dt,-2.2*\yspac) and (4.8*\Dt,-2.2*\yspac) .. node[below,blue!50!red]{Particle momenta push} (5*\Dt,-1.4*\yspac);
\draw[->,blue!50!red,>=stealth,thick] (4.17*\Dt,0.5*\yspac) -- node[anchor=west,blue!50!red]{Field gathering} (4.17*\Dt,-1.2*\yspac);
% c) Maxwell solver
\draw[<-,blue,>=stealth,thick] (4.2*\Dt,1.5*\yspac) -- node[anchor=east,blue]{Inverse FFT}(4.25*\Dt,2.5*\yspac);
\draw[->,blue,>=stealth,thick] (2.2*\Dt,3.5*\yspac) .. controls
(2.4*\Dt,4.5*\yspac) and  (3.6*\Dt,4.5*\yspac)
.. node[below,blue]{Galilean PSATD} (3.9*\Dt,3.5*\yspac);
\draw[->,blue,>=stealth,thick] (2.2*\Dt,3.5*\yspac) .. controls
(2.4*\Dt,4.8*\yspac) and  (3.8*\Dt,4.8*\yspac)
.. node[above,blue]{Eqs. (\ref{eq:update_eq_E}) and (\ref{eq:update_eq_B})} (4.3*\Dt,3.5*\yspac);
\end{tikzpicture}
\caption{\textbf{Illustration of the field push and particle momenta push, in the averaged Galilean PSATD algorithm.} The quantities represented
in black are the ones that are known just before the field and particle push. (Note that this includes the deposited charge and current $\spectral{\rho}$ and $\spectral{J}$.) The quantities in blue and purple are the ones that are being computed during the field and particle push. As part of the field push (blue arrows), the regular fields $\spectral{E}$ and $\spectral{B}$ are updated, and the averaged fields $\langle \spectral{E} \rangle$ and $\langle \spectral{B} \rangle$ are calculated and transformed to the spatial grid. As part of the particle push, the averaged fields $\langle E \rangle$ and $\langle B \rangle$ are gathered onto the macroparticles, in order to update the macroparticles' momenta. The rest of the PIC cycle (e.g. charge and current deposition, particle position push) is not shown here but is identical to the standard Galilean PSATD \cite{LehePRE2016,KirchenPoP2016}.
See the Methods section for more details on the PIC loop and exact definitions of the notation.}
\label{fig:pic_loop}
\end{figure}
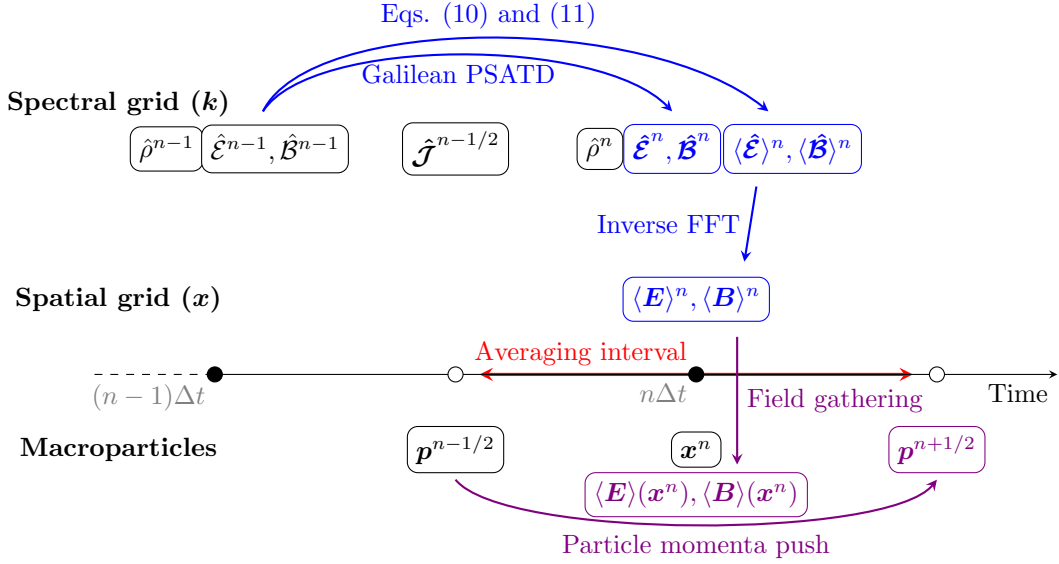

We note that with most Maxwell solvers (for example, the FDTD and PSTD algorithms), the time evolution of the electromagnetic fields \emph{within} one time step is in general not known. However, this evolution is indeed known in the case of the PSATD algorithm. More specifically, as part of the derivation of the PSATD algorithm \cite{Haber,VayJCP2013,LehePRE2016}, the time evolution of the $\vec{E}$ and $\vec{B}$ fields in Fourier space is calculated analytically. Here, we propose to average this analytical expression over one timestep in Fourier space (see equations (\ref{eq:update_eq_E}) and (\ref{eq:update_eq_B}) in the Methods section), and then to transform these averaged fields $\langle\vec{E}\rangle$ and $\langle\vec{B}\rangle$ to real space, where they are gathered onto the macroparticles and then discarded. (However, the \emph{unaveraged} $\vec{E}$ and $\vec{B}$ fields are still kept in memory, and are updated by the standard Galilean PSATD equations \cite{LehePRE2016,KirchenPoP2016} at each PIC iteration.) The corresponding modified PIC loop is illustrated in Fig.~\ref{fig:pic_loop} and described in more detail in the Methods section.

In the rest of this article, we refer to this new scheme as the \emph{averaged Galilean PSATD algorithm}, since it combines the Galilean PSATD scheme \cite{LehePRE2016,KirchenPoP2016,KirchenPRE2020} and the temporal average of the fields over one timestep. By construction, the averaged Galilean PSATD algorithm inherits the main advantages of the Galilean PSATD scheme: it has a low amount of spurious numerical dispersion (for high-order spatial derivatives \cite{VincentiCPC2016,JalasPoP2017}) and does not have a CFL limit. In addition, as shown in the next sections, the averaged Galilean PSATD algorithm efficiently mitigates the NCI for large timesteps.

\subsection*{Stability analysis for a uniform plasma drifting at relativistic velocity}

In order to analyze the stability of the new averaged Galilean PSATD algorithm, we consider again the case of a two-dimensional uniform plasma drifting with a relativistic velocity. We derived the theoretical dispersion equation for this system, by using a similar method as for the standard Galilean PSATD algorithm \cite{LehePRE2016} -- while taking into account the additional average in time. The full derivation of this theoretical dispersion equation is given in the Supplementary Information. By solving this dispersion equation numerically, we can extract the NCI growth rate $Im(\omega)/\omega_{p,r}$ as a function of $\vec{k}$. In addition, we also performed actual PIC simulations for the same system. We used the PIC code WarpX, in which we implemented the averaged Galilean PSATD algorithm, and we then extracted the NCI growth rate of the NCI in post-processing.

\begin{figure}[htb!]
\centering
\begin{subfigure}{\linewidth}
\includegraphics[width=\linewidth]{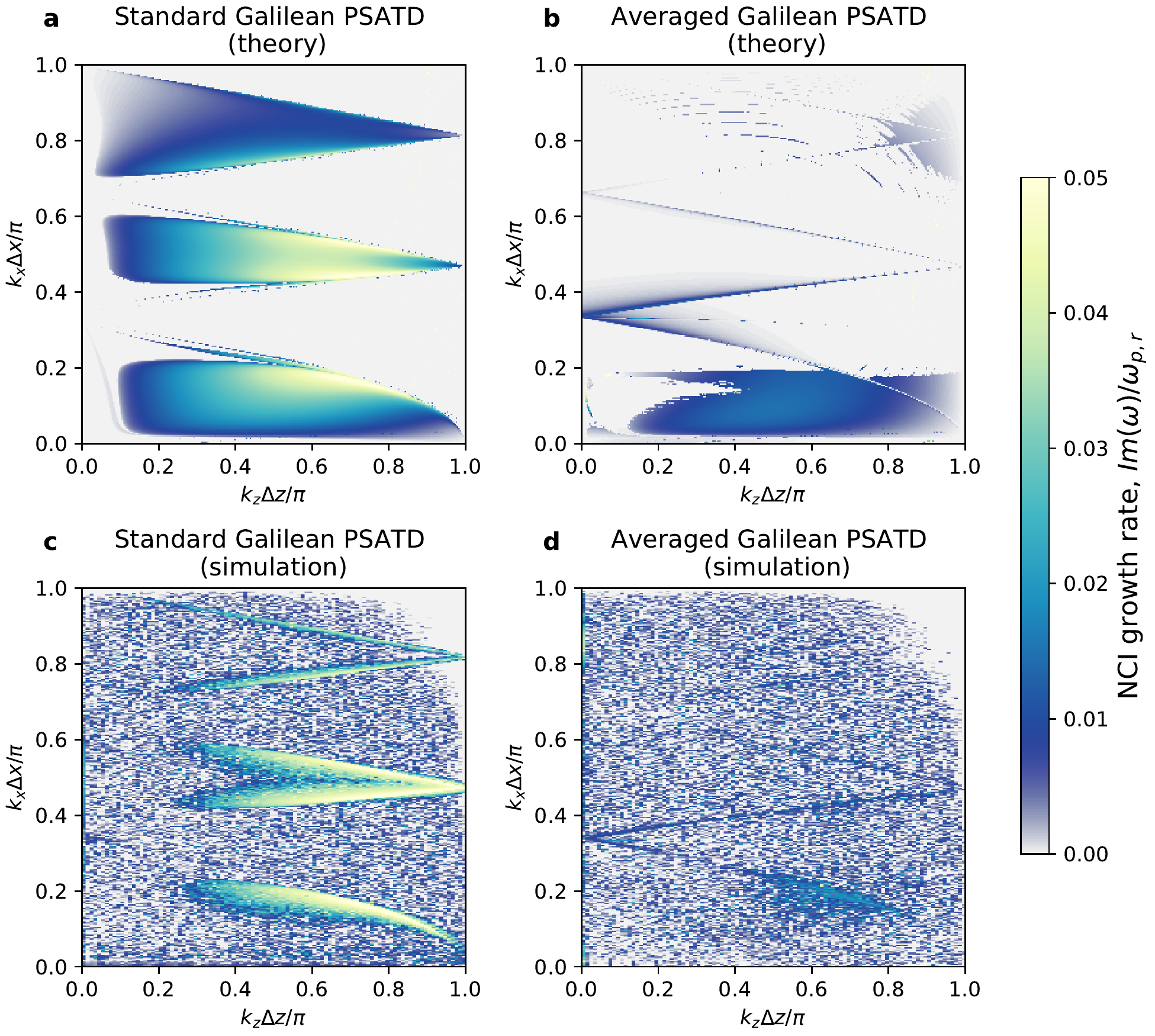}
\end{subfigure}
\caption{\textbf{NCI growth rate: Galilean PSATD vs. averaged Galilean PSATD schemes.} Normalized NCI growth rate  $\text{Im}(\omega) / \omega_{p,r}$ in spectral ($k_x$,$k_z$) space, calculated from the analytical stability analysis (a)-(b) and from WarpX simulation results (c)-(d), obtained using the Galilean PSATD and averaged Galilean PSATD schemes, at infinite spectral order, with large time step $c\Delta t = \Delta z=6 \Delta x$ and slightly detuned Galilean velocity $v_{gal} = 0.99 \, v_0$.}
\label{fig:nci_growth_rate_theory_vs_wx}
\end{figure}

The growth rates extracted from both the WarpX simulations and the theoretical dispersion equation are shown in Fig.~\ref{fig:nci_growth_rate_theory_vs_wx} - both for standard Galilean PSATD (left panels) and for the averaged Galilean PSATD (right panels). For this case, we used the same parameters as for the lower right panel of Fig.~\ref{fig:nci_growth_rate_gal}, \textsl{i.e.}, $\gamma_b = 130$, $\Delta x = 6.4 \times 10^{-2}\, k_{p,r}^{-1}$, a large timestep $c\Delta t = \Delta z= 6\Delta x$, and a detuned velocity $v_{gal} = 0.99\,v_0$. (Recall from the previous sections that the case of a detuned velocity is the one for which using a large timestep presents a major issue.)

As can be seen in Fig.~\ref{fig:nci_growth_rate_theory_vs_wx}, the theoretical
predictions (upper panels) and simulation results (lower panels) are in good
agreement, which confirms that the theoretical dispersion equation correctly
captures the nature of the instability. (Note that growth rate measured from
simulations is typically noisy, which limits the comparison.) More importantly,
both the theoretical predictions and simulation results show that averaged
Galilean PSATD (right panels) strongly reduces the growth of the instability
compared to the standard Galilean PSATD (left panels). This confirms that
averaging the fields in time inhibits spurious resonances
with under-resolved electromagnetic modes, and thereby enables stable
simulations with large timesteps.

\subsection*{Application to simulations of plasma wakefield in a Lorentz-boosted frame}
\label{sec:LWFA}

\begin{figure}[htb!]
\centering
\begin{subfigure}{\linewidth}
\includegraphics[width=\linewidth]{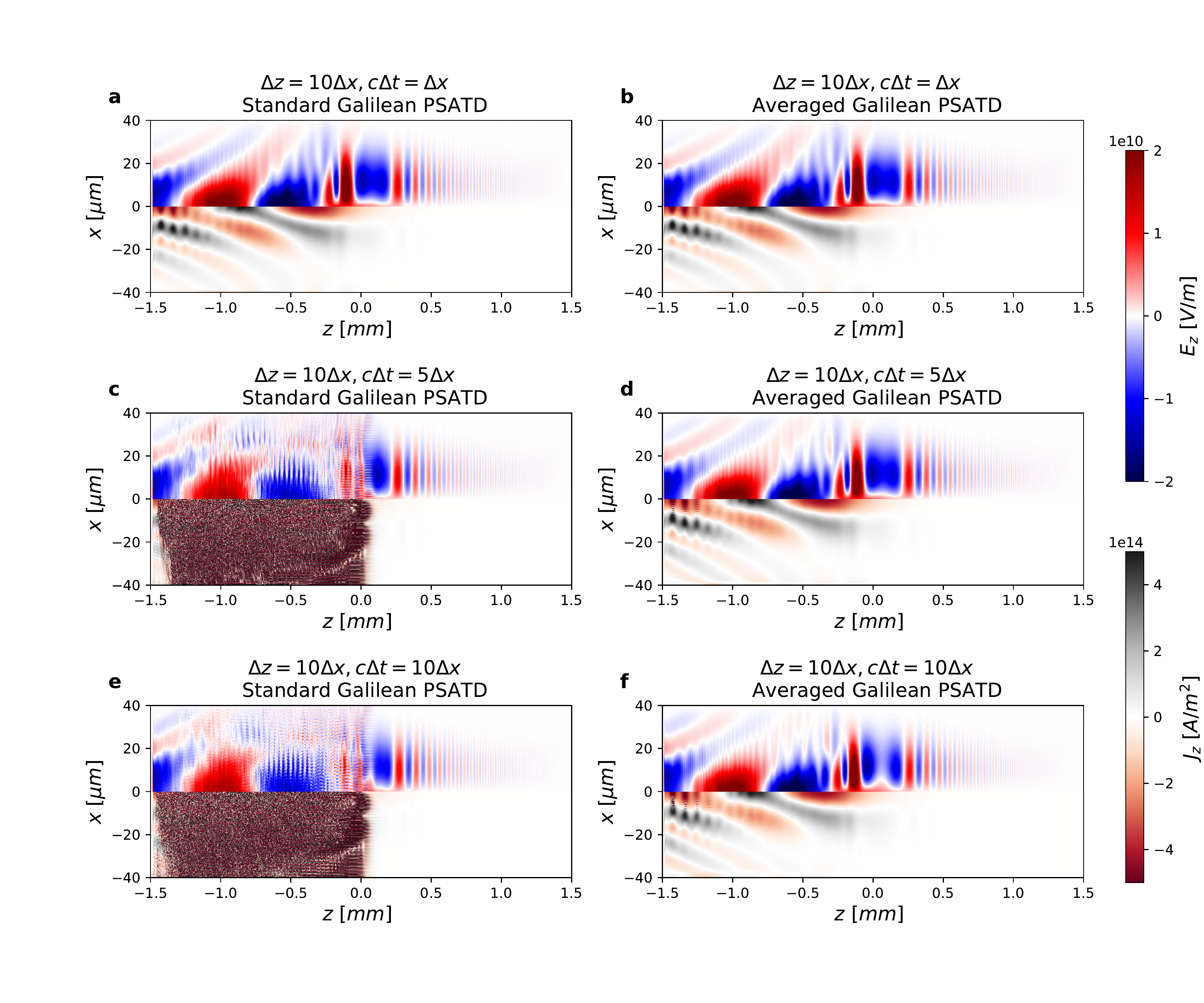}
\end{subfigure}
\caption{\textbf{Instability mitigation in a 2D laser-wakefield simulation with large timestep}. The upper half of each subplot shows the $E_z$ field seen by the macro-particles (\textsl{i.e.}, the regular $E_z$ field for the standard Galilean PSATD, and the averaged $\langle E_z \rangle$ field for the averaged Galilean PSATD). The lower half of each subplot shows the longitudinal current density $J_z$ of the plasma. All the fields are shown in the boosted frame. The different subplots correspond to the standard Galilean PSATD (a,c,e) and averaged Galilean PSATD (b,d,f), with $c\Delta t = \Delta x$ (a-b), $c\Delta t =0.5 \Delta z  = 5 \Delta x$ (c-d) and $c\Delta t = \Delta z  = 10 \Delta x$ (e-f).}
\label {fig:lwfa_simulation}
\end{figure}

This section illustrates that the stability properties observed in the case of a uniform
plasma also apply to realistic simulations of plasma-based acceleration. To this end,
we first perform two-dimensional (2D) simulations of a laser-wakefield accelerator. In
these simulations, an $x$-polarized Gaussian laser pulse with amplitude $a_0 = 1$,
duration $\tau = 20\,\mathrm{fs}$ and waist $w_0 = 15 \,\mathrm{\mu m}$ propagates
in a matched parabolic plasma channel with a background density of $1.0 \times 10^{18}
\,\mathrm{cm}^{-3}$. The simulation runs in a Lorentz-boosted frame ($\gamma_{b}= 30$)
with a nodal PSATD solver with finite order 16 along the $x$ direction and 32 along
the $z$ direction \cite{VincentiCPC2016,JalasPoP2017,KirchenPRE2020}. The longitudinal
resolution (in the boosted frame) is set to
$\Delta z = 2\gamma_b \lambda_{lab} /32 = 1.52 \,\mathrm{\mu m}$, while the transverse
resolution is $\Delta x = 0.15\,\mathrm{\mu m}$,
so that $\Delta z = 10 \Delta x$. We run the simulation with the standard and averaged Galilean
PSATD, and with a small timestep ($c\Delta t = \Delta x$) as well as large timesteps
($c\Delta t = 5\Delta x$ and $c\Delta t = 10 \Delta z$). Fig.~\ref{fig:lwfa_simulation})
displays snapshots of the longitudinal electric field $E_z$ and of the longitudinal current density $J_z$ \emph{in the boosted frame}, for these different cases.
In these colormaps, the rapid oscillations of $E_z$ for $z > 0\,\mathrm{mm}$ correspond to the
longitudinal component of the laser field, which undergoes significant non-linear evolution and red-shifting,
while the slow oscillations of $E_z$ and $J_z$ for $z < 0\,\mathrm{mm}$ correspond to the plasma wakefield.

As expected, the standard Galilean PSATD is stable
for a small timestep (panel (a)), but unstable for large timesteps (panels (c) and (e)). More specifically,
in panels (c) and (e), spurious oscillations rapidly grow in the wakefield and severely disrupt its structure.
We also note that the results of the averaged Galilean PSATD with a small timestep (panel (b))
are almost indistinguishable from those of the standard Galilean PSATD (panel (a))
- thereby confirming that averaging the fields in time preserves the essential physics.
More importantly, for large timestep (panel (d) and panel (f)), the averaged Galilean PSATD achieves
stability while preserving the overall structure of the wakefield. Indeed for $c\Delta t = 5\Delta x$ (panel (d))
the $E_z$ field is still almost indistinguishable from that of panel (a). For $c\Delta t = 10\Delta x$ (panel (f)),
small differences become noticeable, especially in the red-shifted laser oscillations - although they
hardly affect the structure of the accelerating wakefield. This may indicate that
this value of $\Delta t$ starts to reach the limit for which the simulation is
not well-resolved in time anymore.

We also note that, both for $c\Delta t = 5\Delta x$ (panel (d)) and $c\Delta t = 10\Delta x$ (panel (f)),
small transverse oscillations become noticeable in $J_z$ for $z < -1.0\,\mathrm{mm}$.
These oscillations may be due to the remaining non-zero growth rate of the averaged Galilean algorithm
(see the growth rates represented in panel (b) and (d) of Fig.~\ref{fig:nci_growth_rate_theory_vs_wx}).
However, their magnitude is small enough that they do not lead to a modulation of the electric field,
hence they do not affect the dynamics. Again, this represents a clear improvement compared to the
standard Galilean algorithm (panel (b) and (e)).

%We first vary $\Delta x$ from $2.43\,\mu m$ ($\Delta z/\Delta x = 1$) to $0.16\,\mu m$
%($\Delta z/\Delta x = 15$) while keeping $c\Delta t = \Delta z$
%(\textsl{i.e.}, large timestep compared to $\Delta x$). For each value of $\Delta x$, we
%run the simulation with the standard Galilean PSATD and averaged Galilean PSATD.

%The corresponding plot is shown in panel (a)
%of Fig.~\ref{fig:lwfa_simulation}. For the standard Galilean algorithm (black curve)
%the simulation is stable for $\Delta z/\Delta x = 1$, but unstable for even moderately
%elongated cells ($\Delta z/\Delta x \approx 3$), due to the fact that we are using
%a large timestep $c\Delta t=\Delta z$. By contrast, the averaged Galilean PSATD
%(red curve) remains stable even for highly elongated cells (e.g. $c\Delta t/\Delta z \sim 10$).

%Next, we choose the case $\Delta z/\Delta x=10$ as a representative example, and
%we examine the integrity of the simulated physics in more detail.
%For this case,

We now go further and show that this algorithm is applicable to a full-scale
three-dimensional (3D) setting, by considering the 3D simulation of charged particle beam-driven wakefield.
In this simulation, a 1 nC Gaussian electron beam propagates in a plasma with a
background density of $1.0 \times 10^{17} \,\mathrm{cm}^{-3}$, and experiences
a typical evolution whereby the head of the beam erodes while the tail of the beam performs
betatron oscillations in the generated wakefield. The electron beam initially has a mean
Lorentz factor $\gamma = 2000$, with a relative RMS spread $\Delta \gamma/\gamma = 0.01$,
and a transverse and longitudinal RMS size of $5 \,\mathrm{\mu m}$ and $20\,\mathrm{\mu m}$
respectively. The simulation is run in a Lorentz-boosted frame ($\gamma_{b}= 5.6$),
with 512$^3$ cells of size $\Delta x = \Delta y = 0.78\,\mathrm{\mu m}$,
$\Delta z = 5\,\mathrm{\mu m} = 6.4 \,\Delta x$ (in the boosted frame),
and a nodal PSATD solver with finite order 16 \cite{VincentiCPC2016,JalasPoP2017,KirchenPRE2020}.
In order to verify again that the averaged Galilean PSATD algorithm preserves the
physics of interest, we run the simulation both with the standard Galilean PSATD algorithm
and a small timestep $c\Delta t = \Delta x$ (fiducial case) and with the averaged
Galilean PSATD algorithm and a large timestep $c\Delta t = \Delta z$.
In both cases, we ran the WarpX code on the Summit supercomputer, using 24 GPUs
with domain decomposition along $z$.

\begin{figure}[htb!]
\centering
\begin{subfigure}{\linewidth}
\includegraphics[width=\linewidth]{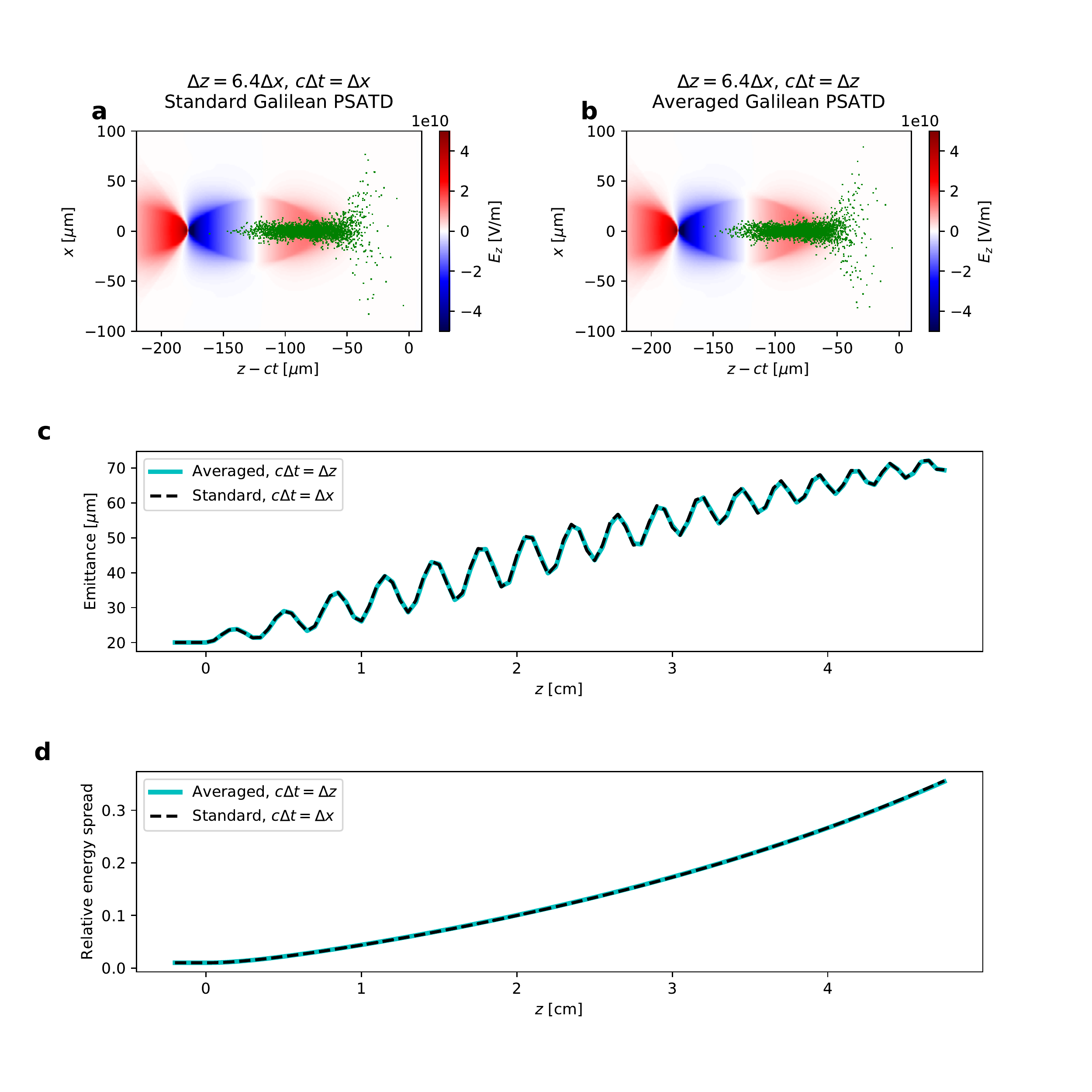}
\end{subfigure}
\caption{\textbf{Algorithm comparison for a 3D plasma-wakefield simulation.}
(a,b) Snapshot of the $E_z$ field ($\langle E_z \rangle$ in the case of the averaged Galilean PSATD) in the laboratory frame,
for the small-timestep standard Galilean PSATD simulation (a)
and large-timestep averaged Galilean PSATD simulation (b).
The green dots are representative random samples of the macroparticles
in the beam driver. (c,d) Evolution of the emittance and relative energy spread
of the beam, in the laboratory frame.}
\label {fig:pwfa_simulation}
\end{figure}

The top panels of Figure~\ref{fig:pwfa_simulation} display colormaps of the
wakefield in the laboratory frame, which were reconstructed on-the-fly during the
boosted-frame simulation. Again, the simulation with the large timestep and
the averaged Galilean PSATD is stable (panel (b)) and the simulated wakefield
is almost indistinguishable from that produced from the fiducial small-timestep
simulation (panel (a)). In addition, panels (c) and (d) in Figure~\ref{fig:pwfa_simulation}
show the evolution of the emittance and relative energy spread of the driver beam
in the laboratory frame, as it undergoes head erosion and betatron oscillation.
This is obtained from laboratory-frame particle data that is reconstructed
on-the-fly during the boosted-frame simulation. As can be seen, the evolution
of these beam quantities show excellent agreement between the fiducial
small-timestep simulation and the large-timestep with the averaged Galilean PSATD.

Thus, in the above example, the averaged Galilean PSATD allowed
stable simulations to be run with a large timestep while preserving the integrity of
the physics at stake. We note that, as a consequence of the large timestep,
the simulation using the averaged Galilean PSATD exhibited a $5\times$ overall speed-up
compared to the small-timestep standard Galilean PSATD simulation, on the Summit supercomputer.

\section*{Discussion}

In this paper, we proposed a modified PIC algorithm that enables stable
boosted-frame simulations of plasma-based acceleration with a large timestep
$c\Delta t \gg \Delta x$, where $\Delta x$ represents the smallest cell size.
This was achieved by using a CFL-free Galilean PSATD solver and by
averaging the $\vec{E}$ and $\vec{B}$ fields in time, in order to inhibit spurious
resonances with under-resolved, aliased electromagnetic modes. We demonstrated
this novel scheme in realistic 2D and 3D plasma-wakefield simulations.

We note that the proposed algorithm could certainly be further refined and improved upon
in the future. For instance, although the proposed algorithm strongly reduces
the NCI growth rate for large timesteps, it does not completely eliminate it.
As a consequence, the NCI at large timesteps could still be an issue for certain
sets of parameters. Similarly, we observed in separate tests that realistic 3D simulations
of laser-driven wakefield could remain unstable in certain cases for large
timesteps, even with the proposed algorithm. %Studies of the remaining instabilities are underway and novel methods that build upon the time-averaged algorithm presented here to mitigate them are being tested and show very promising results, to be presented in another publication.

In conclusion, this work demonstrates that it is
possible to run boosted-frame simulations with a much larger timestep than
the traditional CFL limit, while still accurately capturing the physics.
This new development enables potential speedups of an order of magnitude or
more, opening up a new area of investigation within the field of first-principles,
Particle-In-Cell modeling of plasma-wakefield particle accelerators,
whereby the simulation timestep is chosen much more freely than before.

Although the present work is focused on simulations of plasma accelerators and on
a particular method (the averaged Galilean PSATD), it could have a wider impact.
For instance, even though the algorithm proposed here builds specifically upon the PSATD framework,
the central idea (namely averaging the fields in time) is fairly general and could thus
also guide the future development of similar solutions for FDTD-based methods.
In addition, beyond the plasma accelerator community, this work may be of interest
to the modeling of advanced light sources concepts, coherent synchrotron radiation
in particle accelerators, astrophysical shocks or beam-plasma instabilities of
astrophysical relativistic jets, which can also utilize the boosted-frame PIC
method for accurate modeling from first principles. We also envision that the
method that is used in this paper can be employed to overcome similar timestep
limitations in PIC simulations that do not employ the boosted-frame technique,
with impact to a much wider range of applications.

\section*{Methods}

While the main text gave a brief overview and discussed the main advantages of the averaged Galilean PSATD algorithm, in this section we present the mathematical details of the proposed scheme, including the derivation of the discretized averaged fields.

\subsection*{Derivation of the averaged fields}
Throughout the paper, we use the notation $\langle \Spectral{F}\rangle^{n} $ to refer to the averaged field within the time interval $[(n-\frac 12)\Delta t, (n+\frac 12)\Delta t]$, defined as
%We first define the averaged field $\langle \Spectral{F}\rangle^{n} $ within the time interval $[(n-\frac 12)\Delta t, (n+\frac 12)\Delta t]$ as
\begin{equation}
\label{eq:average}
\langle \Spectral{F}\rangle ^{n} = \frac{1}{\Delta t}\int_{(n-1/2)\Delta t }^{(n+1/2)\Delta t} \Spectral{F}(\vec{k}, \tau) d\tau \,, %\quad \forall t \in [(n-1/2)\Delta t, (n+1/2)\Delta t].
\end{equation}
where $\Spectral{F} = \Spectral{F}(\vec k, t) = \int{ \vec F(\vec{x},t) e^{-i\vec k \cdot \vec x} d^3\vec{x}}$ refers to the Fourier transform of the $\vec{F}(\vec x, t)$ field.

% We remark, since the novel scheme is based on the Galilean PSATD algorithm, originally introduced in \cite{LehePRE2016}, for convenience we will use the notations used in \cite{LehePRE2016} as much as possible.

In the Galilean coordinates drifting at $\vec{v}_{gal}$, the $p$-order discretized
Maxwell equations transformed to Fourier space read\cite{KirchenPRE2020}:
\begin{align}
\left ( \frac{\partial}{\partial t} - i[\vec{k}] \cdot \vec{v}_{gal} \right)^2\Spectral{B} + c^2[k]^2 \Spectral{B} &= \frac{1}{\varepsilon_0} i [\vec{k}] \times \Spectral{J},
\\
\left ( \frac{\partial}{\partial t} - i[\vec{k}] \cdot \vec{v}_{gal} \right) ^2\Spectral{E} + c^2[k]^2 \Spectral{E} &= -\frac{c^2}{\varepsilon_0} \spectral{\rho}i[\vec{k}] - \frac{1}{\varepsilon_0} \left(\frac{\partial}{\partial t} - i[\vec{k}]\cdot\vec{v_{gal}}\right)\Spectral{J},
\end{align}
where $[k] =\sqrt {[\vec k] ^2}= \sqrt{[k_x]^2 + [k_y]^2 + [k_z]^2}$ and where $[k_u]$ with $u=\{x, y, z\}$ is the Fourier transform of the $p$-order discretized stencil $\hat{\nabla}_u$
(\textsl{i.e.}, such that a $p$-order Taylor expansion yields $\hat{\nabla}_u F = \partial_u F + O(\Delta u^p)$.) \cite{VincentiCPC2016,JalasPoP2017}.

As explained in \cite{LehePRE2016}, these equations can be integrated analytically under the assumption that the current $\Spectral{J}$ is constant over one timestep, and that the fields $\Spectral{E}, \Spectral{B}, \Spectral{J}$ and $\spectral{\rho}$ satisfy the conservation equations.
More specifically, assuming that the $\vec{E}$ and $\vec{B}$ fields are known at $t=(n-1)\Delta t$, and under the assumption that $\Spectral{J}(\vec{k}, t)$ is constant and equal to $\Spectral{J}^{n-1/2}(\vec k)$ over the time interval $[(n-1)\Delta t, n\Delta t]$, we can obtain the expressions of $\Spectral{E}(\vec{k}, t)$, $\Spectral{B}(\vec{k}, t)$ as a function of $\Spectral{E}^{n-1}$, $\Spectral{B}^{n-1}$, $\Spectral{J}^{n-1/2}$, $\Spectral{\rho}^{n-1}$, $\Spectral{\rho}^n$:
\begin{align}
\begin{split}
\label{eq:analytical_eq_E}
\Spectral{E}(\vec{k}, t) &= \left[ \Spectral{E}^{n-1} - \frac{\vec{\alpha}_1}{c^2[k]^2(1-\nu^2)} - \frac{\vec{\beta}_1}{c^2[k]^2} \right] \cos{[c[k](t-(n-1)\Delta t)]}e^{i\nu c[k](t-(n-1)\Delta t)} \\
& + \frac{\vec{\alpha}_1}{c^2[k]^2(1-\nu^2)} + \frac{\vec{\beta}_1}{c^2[k]^2}e^{i\nu c[k](t-(n-1)\Delta t)} \\
& + \frac{1}{c[k]}\left[c^2 i \vec{[k]} \times \Spectral{B}^{n-1} - \frac{1}{\varepsilon_0}\Spectral{J}^{n-1/2}+ i\nu \frac{\vec{\alpha}_1}{c[k](1-\nu^2)} \right]\sin{[c[k](t-(n-1)\Delta t)]}e^{i\nu c[k](t-(n-1)\Delta t)} \\
\end{split} \\[5pt]
\begin{split}
\label{eq:analytical_eq_B}
\Spectral{B}(\vec{k}, t) &= \left[ \Spectral{B}^{n-1} - \frac{\vec{\alpha}_2}{c^2[k]^2(1-\nu^2)} \right] \cos{[c[k](t-(n-1)\Delta t)]}e^{i\nu c[k](t-(n-1)\Delta t)} + \frac{\vec{\alpha}_2}{c^2[k]^2(1-\nu^2)} \\
&+ \frac{1}{c[k]}\left[ - i \vec{[k]} \times \Spectral{E}^{n-1} + i\nu  \frac{\vec{\alpha}_2}{c[k](1-\nu^2)}  \right]\sin{[c[k](t-(n-1)\Delta t)]}e^{i\nu c[k](t-(n-1)\Delta t)}
\end{split}
\end{align}
where $\nu = \vec{[k]} \cdot \vec{v_{gal}}/c[k]$ and
\begin{align}
\vec{\alpha_1} &= \frac{i\nu c[k]}{\varepsilon_0}\Spectral{J}^{n-1/2} - \frac{c^2}{\varepsilon_0}\frac{\spectral{\rho}^{n} - \spectral{\rho}^{n-1}e^{i \nu c[k] \Delta t}}{1-e^{i\nu c[k]\Delta t}}i\vec{[k]}\\
\vec{\beta_1} &= \frac{c^2}{\varepsilon_0}  \frac{\spectral{\rho}^{n} - \spectral{\rho}^{n-1}}{1-e^{i \nu c[k] \Delta t}}i\vec{[k]} \\
\vec{\alpha_2} &= \frac{1}{\varepsilon_0}i\vec{[k]} \times \Spectral{J}^{n-1/2}
\end{align}

Strictly speaking, these expressions of $\Spectral{E}(\vec{k}, t)$, $\Spectral{B}(\vec{k}, t)$ in
Eqns. (\ref{eq:analytical_eq_E}) and (\ref{eq:analytical_eq_B}) are only valid for $t$ in the interval $[(n-1)\Delta t, n\Delta t]$ (because of the assumption $\Spectral{J}(\vec{k}, t) = \Spectral{J}^{n-1/2}(\vec k)$). However, we assume that there are also approximately
valid on the interval $[(n-1/2)\Delta t, (n+1/2)\Delta t]$ over which the fields are
averaged (see Fig.~\ref{fig:pic_loop}). This is valid if $\Spectral{J}$ varies slowly from one timestep to the next - \textsl{i.e.}, if the plasma response is well-resolved in time.

Under these assumptions, we average Eqns. (\ref{eq:analytical_eq_E}) and (\ref{eq:analytical_eq_B}) in time as defined in Eq.~(\ref{eq:average}), and obtain:
\begin{align}
\begin{split}
\label{eq:update_eq_E}
\langle \Spectral{E} \rangle^{n}& = \Psi_1 \Spectral{E}^{n-1}  -  ic \Psi_2  \frac{\vec {[k]}}{[k]} \times  \Spectral{B}^{n-1} +  \Big( \frac{i \nu c[k]}{\varepsilon_0} A_1 + \frac{\Psi_2}{c[k]\varepsilon_0} \Big) \Spectral{J}^{n-1/2} \\
& \quad + C_{\rho}  (A_2-A_1)  \spectral{\rho}^{n}  \frac{\vec{[k]}}{[k]}+ C_{\rho}  (\theta^2 A_1-A_2)  \spectral{\rho}^{n-1} \frac{\vec{[k]}}{[k]} \,,
\end{split} \\[5pt]
\label{eq:update_eq_B}
\langle \Spectral{B} \rangle^{n} & = \Psi_1 \Spectral{B}^{n-1} +  \frac{i}{c}  \Psi_2  \frac{\vec{[k]}}{[k]} \times  \Spectral{E}^{n-1}+ \frac{i[k]}{\varepsilon_0} A_1 \frac{\vec{[k]}}{[k]} \times \Spectral{J}^{n-1/2}.
\end{align}
Here again, $\nu = \vec{[k]} \cdot \vec{v_{gal}}/c[k]$,  $\theta = e^{i\vec{[k]} \cdot \vec{v_{gal}} \Delta t/2}$, and the other coefficients are given by:
\begin{subequations}
\begin{align}
& \ C_{\rho} = \frac{i c^2 [k]}{\varepsilon_0 (1 - \theta^2)} \,, \\
& A_1 = \frac{\Psi_1 -1 + i \nu \Psi_2}{c^2 [k]^2 (\nu^2 -1)} \,, \\
& A_2 = \frac{\Psi_3 - \Psi_1}{c^2 [k]^2} \,, \\
& \Psi_1 = \theta \frac{(S_1 + i\nu C_1) - \theta^2 (S_3+i \nu C_3)}{c[k]\Delta t(\nu^2-1)} \,, \\
& \Psi_2 =\theta \frac{(C_1 - i\nu S_1) - \theta^2 (C_3 - i \nu S_3)}{c[k]\Delta t(\nu^2-1)} \,, \\
& \Psi_3 = \frac{i \theta (1 - \theta^2)}{c[k] \Delta t \nu} \,,
\end{align}
\end{subequations}
with $C_m = \cos(m \, c [k] \, \Delta t/2)$ and $S_m = \sin(m \, c [k] \, \Delta t/2)$ for $m = 1,2,3$.

\subsection*{PIC cycle overview}

Fig.~\ref{fig:pic_loop} gives an overview of a key part of the PIC loop for the averaged Galilean PSATD algorithm. Here we describe the exact PIC loop in more detail. Assuming that we originally
know the particle positions and momenta $\vec{x}^n$ at $\vec{p}^{n-1/2}$ and the fields $\vec{E}^{n-1}$ and $\vec{B}^{n-1}$, the loop consists of the following steps:
\begin{enumerate}
\item Deposit the charge and current densities of the particles onto the spatial grid. In particular, we deposit the charge density $\rho^{n}$ at time $t = n \Delta t$ from the particle positions $\vec{x}^{n}$ and the current density $\vec{J}^{n-1/2}$ at time $t = (n-\frac 12) \Delta t$ from the particle positions $\vec{x}^{n-1}$ and $\vec{x}^{n}$ and the particle velocities $\vec{v}^{n-1/2}$;
\item Transform all relevant physical quantities from physical space to Fourier space;
\item Compute the new electromagnetic fields in Fourier space $\Spectral{E}^{n}$ and $\Spectral{B}^{n}$, from the charge and current densities $\spectral{\rho}^{n-1}$ (available from the previous PIC iteration), $\spectral{\rho}^{n}$ and $\Spectral{J}^{n-1/2}$;
\item Compute the averaged electromagnetic fields, $\langle \Spectral{E} \rangle^n$ and $\langle \Spectral{B} \rangle^n$;
% , by taking the analytical expressions of $\Spectral{E}$ and $\Spectral{B}$, theoretically valid only within the time interval $[(n-1) \Delta t, n \Delta t]$ and used to carry out the computation in point 4. above, and by averaging them over the time interval $[(n-\frac 12) \Delta t, (n+\frac 12) \Delta t]$. This is based on the assumption that the current density remains constant and known also within the extra half time interval $[n \Delta t, (n+\frac 12) \Delta t]$;
\item Transform all relevant physical quantities from Fourier space back to physical space;
\item Gather the averaged electromagnetic fields, in physical space, from the spatial grid onto the particles;
\item Push the particles by updating their positions and momenta based on the current values of the averaged electromagnetic fields, $\langle \vec{E} \rangle^n$ and $\langle \vec{B} \rangle^n$, whose precise definition is given in the next section. In particular, the momenta are updated from $\vec{p}^{n-1/2}$ to $\vec{p}^{n+1/2}$ and the positions are then updated from $\vec{x}^{n}$ to  $\vec{x}^{n+1}$
\end{enumerate}
\subsection*{WarpX code}
%Copy-pasted, needs to be updated
We implemented the averaged Galilean PSATD algorithm of arbitrary spectral order in the code WarpX \cite{warpx}  in Cartesian coordinates.
WarpX is an open-source PIC code that combines advanced algorithms with adaptive mesh refinement to allow challenging simulations of a multi-stage plasma-based TeV acceleration relevant for future high-energy physics discoveries. WarpX relies on the ECP AMReX co-design center for mesh refinement and runs on CPU and GPU-accelerated platforms, and production simulations have run on Cori KNL at NERSC and on Summit at OLCF.

\section*{Code availability}
WarpX code is available at \url{https://github.com/ECP-WarpX/WarpX}.

\bibliographystyle{unsrt}
\bibliography{references}

\section*{Acknowledgments}
The authors acknowledge the discussions and code contributions from
the whole WarpX development team, that made this work possible. We are also thankful to Andrew Myers for valuable feedback on the manuscript.
This research used resources of the Oak Ridge Leadership Computing Facility
at the Oak Ridge National Laboratory, which is supported by the Office of
Science of the U.S. Department of Energy under Contract No. DE-AC05-00OR22725.

\section*{Author contributions}
O.~S. derived the equations of the averaged Galilean PSATD algorithm,
implemented them in the WarpX code, and extensively tested the algorithm.
R.~L. proposed the initial idea of the averaged Galilean PSATD algorithm
and provided guidance for the derivation and implementation of the algorithm.
J.~V. initiated the research into an algorithm that enables larger time steps,
helped the implementation and testing of an early prototype of the new
algorithm in the Warp code and provided guidance to the overall project and paper writing.
M.~T. and Y.~Z. performed additional simulations, and E.~Z. and M.~T. made major
contributions to the implementation in WarpX.
O.~S. and R.~L. wrote the paper, with key inputs from M.~T., E.~Z., Y.~Z. and J.~V.

\section*{Funding}
This research was supported by the Exascale Computing Project (No. 17-SC-20-SC),
a collaborative effort of the U.S Department of Energy Office of Science and the
National Nuclear Security Administration, as well as by the Director, Office
of Science, Office of High Energy Physics, U.S. Department of Energy under
Contract No. DEAC02-05CH11231.

\section*{Competing interests}
The authors declare no competing interests.

\section*{Additional information}
The derivation of the dispersion relation for the averaged Galilean
PSATD algorithm is given in Supplementary Information.

\section*{Supplementary information: Derivation of the Dispersion Relation for the Averaged Galilean PSATD Algorithm}
Here we derive the 2D dispersion relation to study the NCI induced by a relativistic plasma flowing through a periodic grid along the $z$-axis with velocity $\vec{v_0} = v_0\vec{u_z}$, where $v_0 = c ( 1 - 1/\gamma_b^2 )^{1/2}$. This is done by combining the discretized Maxwell equations in the Galilean frame and the discretized Vlasov equation, and assuming small perturbations for the electromagnetic fields $\vec{E}, \vec{B}$ and the distribution function $\delta f$. Because the Vlasov equation involves the averaged fields $\langle \vec{E}\rangle$, $\langle \vec{B}\rangle$, we include their expression as a function of the regular fields $\vec{E}$, $\vec{B}$ in the system of equations. Hence, when expressed in spectral space, the different equations of the system are:

\begin{itemize}

\item the discretized Maxwell equations at $n$-th time step in time-symmetrical form \cite{LehePRE2016}:
\begin{align}
\label{eq:supp-maxwellE}
\theta^* c \Spectral{B}^{n} - \theta c \Spectral{B}^{n-1} & = -t_{ck} \frac{i\vec{k}\times (\theta^* \Spectral{E}^{n} + \theta \Spectral{E}^{n-1})}{k} + 2 \chi_4^{'} \frac{\hat{T}}{\varepsilon_0 ck} \frac{\vec{k} \times \Spectral{J}_d^{n-1/2}}{k} \,, \\[5pt]
\begin{split}
\label{eq:supp-maxwellB}
\theta^* \Spectral{E}^{n} - \theta \Spectral{E}^{n-1} & = t_{ck} \frac{i\vec{k}\times (\theta^* c \Spectral{B}^{n} + \theta c \Spectral{B}^{n-1})}{k} - \frac{\hat{T} i\vec{k}}{\varepsilon_0 k^2} (\theta^* \spectral{\rho}^{n} - \theta \spectral{\rho}^{n-1}) \\
& \quad - 2 \chi_4 \frac{\hat{T}}{\varepsilon_0 ck} \left[  \Spectral{J}_d^{n-1/2} - \frac{(\vec{k} \cdot \Spectral{J}_d^{n-1/2} )\vec{k}}{k^2} \right] \,;
\end{split}
\end{align}

\item the perturbed Vlasov equation \cite{LehePRE2016}. Note that, here, we replaced the regular fields $\Spectral{E}, \Spectral{B}$ by the averaged fields $\langle\Spectral{E}\rangle, \langle\Spectral{B}\rangle$ in order to take into account the changes associated with the averaged Galilean PSATD.
\begin{align}
\label{eq:supp-vlasov}
\delta \hat{f}^{n+1/2}(\vec{k_m},\vec{p})&e^{i\vec{k_m}(\vec{v}- \vec{v_{gal}})\Delta t/2}-\hat{f}^{n-1/2}(\vec{k_m},\vec{p})e^{-i\vec{k_m}(\vec{v}-\vec{v_{gal}})\Delta t/2} \nonumber \\
& +q\Delta t \hat{S}(\vec{k_m})\Big[\langle \Spectral{E}^n(\vec{k}) \rangle + \vec{v} \times \langle \Spectral{B}^n(\vec{k}) \rangle \Big] \cdot \frac{\partial f_0}{\partial \vec{p}} = 0 \,;
\end{align}

\item the expression of the averaged field for $t \in [(n-1/2)\Delta t,(n+1/2)\Delta t]$:
\begin{align}
\label{eq:supp-avg}
\langle \Spectral{B}^{n}(\vec{k}, t) \rangle & = \Psi_1 \Spectral{B}^{n-1} +  \frac{i}{c}  \Psi_2  \frac{\vec{k}}{k} \times  \Spectral{E}^{n-1}+ \frac{ik}{\varepsilon_0} A_1 \frac{\vec{k}}{k} \times \Spectral{J}^{n-1/2} \,, \\[5pt]
\begin{split}
\langle \Spectral{E}^{n}(\vec{k}, t) \rangle & = \Psi_1 \Spectral{E}^{n-1}  -  ic \Psi_2  \frac{\vec {k}}{k} \times  \Spectral{B}^{n-1} +  \left( \frac{i \nu ck}{\varepsilon_0} A_1 + \frac{\Psi_2}{ck\varepsilon_0} \right) \Spectral{J}^{n-1/2} \\
& \quad + C_{\rho}  (A_2-A_1)  \spectral{\rho}^{n}  \frac{\vec{k}}{k}+ C_{\rho}  (\theta^2 A_1-A_2)  \spectral{\rho}^{n-1} \frac{\vec{k}}{k} \,.
\end{split}
\end{align}

Here, $\hat{T} = \prod_{i} \big[ 1-\sin(k_i \Delta i/2) \big]$ represents a one-pass binomial smoother, and $\hat{S}(\vec{k_m})$ is the particle shape factor, with  $\vec{k_m} = \vec k+\vec{K_m}$ $(\vec{K_m} =  2\pi\sum_{i}u_i m_i/\Delta i )$ for $\ i =\big\{ x,y,z\big\}$.

As follows from the discrete continuity equation, the corrected current $\Spectral{J}^{n-1/2}$ satisfies
\begin{gather}
\label{eq:supp-continuity}
\Spectral{J}^{n-1/2} =\Spectral{J}_d^{n-1/2}  - \frac{(\vec{k} \cdot \Spectral{J}_d^{n-1/2}) \vec{k}}{k^2} + \frac{(\vec{k} \cdot \vec{v_{gal}}) \vec{k}}{k^2} \frac{\spectral{\rho}^n - \spectral{\rho}^{n-1}\theta^2}{1-\theta^2} \,.
\end{gather}

As in \cite{LehePRE2016}, we use the following Ansatz for the electromagnetic modes:
\begin{subequations}
\label{eq:supp-modes}
\begin{align}
\Spectral{E}^n(\vec{k}) & = \Spectral{E}(\vec{k}) e^{-i(\omega - \vec{k} \cdot\vec{v_{gal}})n\Delta t} \,, \\
\langle\Spectral{E}^n(\vec{k}) \rangle &= \langle\Spectral{E}(\vec{k}) \rangle e^{-i(\omega - \vec{k}\cdot \vec{v_{gal}})n\Delta t} \,, \\
\delta \hat{f}^{n-1/2}(\vec{k_m},\vec{p}) & =\delta \hat{f}(\vec{k_m},\vec{p}) e^{-i(\omega - \vec{k}\cdot \vec{v_{gal}})(n-1/2)\Delta t} \,, \\
\Spectral{ J}_d^{n-1/2}(\vec{k}) & = \Spectral{J}_d(\vec{k}) e^{-i(\omega - \vec{k}\cdot \vec{v_{gal}})(n-1/2)\Delta t} \,, \\
\spectral{\rho}^{n}(\vec{k}) & = \spectral{\rho}(\vec{k}) e^{-i(\omega - \vec{k}\cdot \vec{v_{gal}})n\Delta t} \,,
\end{align}
\end{subequations}
and after some amount of algebra, we derive the following equations for $\Spectral{J}_d(\vec{k})$ and $\spectral{\rho}(\vec{k})$ from the Vlasov equation (see a similar derivation in \cite{LehePRE2016}):
\begin{align}
\label{eq:supp-J}
\Spectral{J}_d & = i\frac{ck \varepsilon_0}{\hat{T}} \left(\xi_1 \langle \Spectral{F} \rangle+(\vec{\xi_2} \cdot \langle \Spectral{F}\rangle)\frac{ \vec{v_0}}{c}\right) \,, \\
\label{eq:supp-rho}
\spectral{\rho} & = \frac{i k\varepsilon_0 } {\hat{T}} (\vec{\xi_3} \cdot \langle \Spectral{F} \rangle) \,,
\end{align}
where $ \langle \Spectral{F}(\vec{k}) \rangle = \langle \Spectral{E} (\vec{k})\rangle + \vec{v_0} \times \langle \Spectral{B}(\vec{k}) \rangle - (\vec{v_0} \cdot  \langle \Spectral{E} (\vec{k})\rangle ) \vec{v_0} / c^2$.

In addition, by substituting the expressions (\ref{eq:supp-continuity}) and (\ref{eq:supp-modes}) into (\ref{eq:supp-maxwellE}), (\ref{eq:supp-maxwellB}), (\ref{eq:supp-avg}), the problem is reduced to the following set of equations to be solved:
\begin{align}
\label{eq:supp-E}
s_{\omega} c\Spectral{B} & = t_{ck} c_{\omega}\frac{ \vec{k} \times \Spectral{E}} {k}+ \chi_4^{'} \frac{\hat{T}}{\varepsilon_0 ck} \frac{i\vec{k} \times \Spectral{J}_d}{k} \,, \\
\label{eq:supp-B}
s_{\omega} \Spectral{E} & = -c_{\omega} t_{ck} \frac{\vec{k} c \Spectral{B}}{k} - i s_{\omega} \frac{\hat{T} \vec{k}}{\varepsilon_0 k^2} \spectral{\rho} -i \chi_4 \frac{\hat{T}}{\varepsilon_0 ck} \left[  \Spectral{J}_d - \frac{(\vec{k} \cdot \Spectral{J}_d )\vec{k}}{k^2} \right] \,, \\
\label{eq:supp-avgE}
\theta^2c \langle \Spectral{B} \rangle & = \Psi_1c \Spectral{B} e^{i\omega \Delta t} + i \Psi_2  \frac{\vec{k}\times  \Spectral{E} }{k}  e^{i\omega \Delta t} + \frac{ickA_1 \hat{T}\theta}{\varepsilon_0}\frac{\vec{k} \times \spectral{J}_d}{k}  e^{\frac{i\omega \Delta t}{2}} \,, \\
\begin{split}
 \label{eq:supp-avgB}
\theta^2 \langle \Spectral{E} \rangle & = \Psi_1 \Spectral{E}e^{i\omega \Delta t} -  ic \Psi_2  \frac{\vec {k}}{k} \times  \Spectral{B}e^{i\omega \Delta t}  +   \frac{ iA_{\nu}\hat{T}} {ck\varepsilon_0}  \theta e^{\frac{i\omega \Delta t}{2}} \left[ \Spectral{J}_d  - \frac{(\vec{k} \cdot \Spectral{J}_d) \vec{k}}{k^2} \right] \\
& \quad +\frac{i \vec{k} \spectral{{\rho} }\hat{T}}{\varepsilon_0 k^2}  \frac{ c^2k^2A_2 (\theta^2-e^{i\omega\Delta t}) + \theta^2 ( c^2k^2 A_1 - \nu A_{\nu} )(e^{i\omega\Delta t}-1)}{ (1-\theta^2)} \,.
\end{split}
\end{align}
\end{itemize}
Here,  $ A_{\nu} = \nu c^2k^2 A_1 - i\Psi_2$, and the ${\xi_1,\vec{\xi}_{2,3}}$ coefficients represent the plasma response (for more details see Appendix A in \cite{LehePRE2016}):
\begin{equation}
\xi_{1} = \frac{\hat{T} \omega_p^2}{\gamma_0 c k} \sum_{m} \frac{\hat{S}^{2}(\vec{k_m})}{s_{\omega}'} \,, \quad
\vec{\xi_{2}} = \frac{\hat{T} \omega_p^2}{\gamma_0 k} \sum_{m} \frac{c_{\omega'}\hat{S}^{2}(\vec{k_m})}{s_{\omega '}^2} \vec{k_m} \,, \quad
\vec{\xi_{3}} = \frac{\hat{T} \omega_p^2}{\gamma_0 k} \sum_{m} \frac{\hat{S}^{2}(\vec{k_m})}{s_{\omega '}^2} \vec{k_m} \,,
\end{equation}
where
\begin{equation}
c_{\omega '} = \cos \left(\frac{\omega - \vec k \cdot\vec v_0 - \vec{K_m} (\vec {v_0} -  \vec v_{gal}) }{2\Delta t^{-1}}\right) \,, \quad
s_{\omega '} = \frac{2}{\Delta t}\sin \left(\frac{\omega - \vec k \cdot\vec v_0 - \vec{K_m} (\vec {v_0} -  \vec v_{gal}) }{2\Delta t^{-1}}\right) \,.
\end{equation}
By projecting equations~(\ref{eq:supp-B}) and (\ref{eq:supp-avgB}) along $y$ and equations~(\ref{eq:supp-J}), (\ref{eq:supp-E}) and (\ref{eq:supp-avgE})  along $x$ and $z$, the final system of equations can be written in the matrix form
\begin{align}
\vec{M_{av}} \vec{U} = 0 \,,
\end{align}
where $\vec{M_{av}}$ is the block matrix
\begin{equation}
% \sbox1{$\begin{matrix}1&0&0\\0&1&0\\0&0&1\end{matrix}$}
% \sbox2{$\begin{matrix}-\theta^2 &0&0\\0&-\theta^2&0\\0&0&-\theta^2\end{matrix}$}
% \sbox0{$\begin{matrix}0&0&0\\0&0&0\\0&0&0\end{matrix}$}
% \vec{M_{av}}=\left[
% \begin{array}{c | c| c }
% \makebox[\wd0]{\large \vec{M}} & \usebox{2}& \makebox[\wd0]{\large \vec{N}} \\
% \hline
% \usebox{0}& \makebox[\wd0]{\large \vec{P}} & \usebox{1} \\
% \hline
% \makebox[\wd0]{{\large \vec{R}}} & \usebox{0}& \makebox[\wd0]{\large \vec{Q}} \\
% \end{array},
% \right]
\vec{M_{av}} =
\begin{bmatrix}
  &         &   & -\theta^2 & 0         & 0         &   &         &   \\
  & \vec{M} &   & 0         & -\theta^2 & 0         &   & \vec{N} &   \\
  &         &   & 0         & 0         & -\theta^2 &   &         &   \\
0 & 0       & 0 &           &           &           & 1 & 0       & 0 \\
0 & 0       & 0 &           & \vec{P}   &           & 0 & 1       & 0 \\
0 & 0       & 0 &           &           &           & 0 & 0       & 1 \\
  &         &   & 0         & 0         & 0         &   &         &   \\
  & \vec{R} &   & 0         & 0         & 0         &   & \vec{Q} &   \\
  &         &   & 0         & 0         & 0         &   &         &
% \sbox1{$\begin{matrix}1&0&0\\0&1&0\\0&0&1\end{matrix}$}
% \sbox2{$\begin{matrix}-\theta^2 &0&0\\0&-\theta^2&0\\0&0&-\theta^2\end{matrix}$}
% \sbox0{$\begin{matrix}0&0&0\\0&0&0\\0&0&0\end{matrix}$}
% \vec{M_{av}}=\left[
% \begin{array}{c | c| c }
% \makebox[\wd0]{\large \vec{M}} & \usebox{2}& \makebox[\wd0]{\large \vec{N}} \\
% \hline
% \usebox{0}& \makebox[\wd0]{\large \vec{P}} & \usebox{1} \\
% \hline
% \makebox[\wd0]{{\large \vec{R}}} & \usebox{0}& \makebox[\wd0]{\large \vec{Q}} \\
\end{bmatrix}
\end{equation}
and $\vec{U}$ is the vector
\begin{align}
\vec{U} = \left( c\spectral{B}_y, \spectral{E}_z,\spectral{E}_x,  c\langle \spectral{B}_y \rangle,  \langle \spectral{E} _z\rangle, \langle \spectral{E}_x \rangle, \frac{\spectral{J}_z}{ck\varepsilon_0}, \frac{\spectral{J}_x}{ck\varepsilon_0}, \frac{\spectral{\rho}}{k\varepsilon_0}\right)^{T} \,.
\end{align}

The resulting dispersion relation is given by the determinant equation
\begin{equation}
\label{eq:supp-det}
\det \boldsymbol{M_{av}} = 0.
\end{equation}

Here, the individual matrices defining $\vec{M_{av}}$ read
\begin{subequations}
\begin{align}
\vec{M} & = e_{\omega}
\begin{bmatrix}
\Psi_1 & - i\Psi_2 k_{xn} & i\Psi_2 k_{zn} \\
-i \Psi_2 k_{xn} & \Psi_1 & 0 \\
 i \Psi_2k_{zn} & 0 &\Psi_1
\end{bmatrix} \,, \\[5pt]
\vec{N} & = \theta \sqrt{e_{\omega}} \, \hat{T}
\begin{bmatrix}
-i A_1  k_{xn} & i A_1 k_{zn} & 0 \\
 i k_{xn}^2  A_{\nu} & - i k_{xn}k_{zn} A_{\nu} & i k_{zn} r_{\omega \nu} \\
-i k_{xn} k_{zn} A_{\nu} & i k_{zn}^2A_{\nu} & i k_{xn} r_{\omega \nu}
\end{bmatrix} \,, \\[5pt]
\vec{P} & =\frac{1}{\hat{T}}
\begin{bmatrix}
i\beta_0^2 \xi_{2x} & -i(1-\beta_0^2) (\beta_0\xi_{2z} + \xi_{1}) & -i\beta_0 \xi_{2x} \\
i\beta_0 \xi_{1} & 0 & -i{\xi_1} \\
i\beta_0 \xi_{3x} & -i(1-\beta_0^2)\xi_{3z} & -i\xi_{3x}
\end{bmatrix} \,, \\[5pt]
\vec{R} & =
\begin{bmatrix}
s_{\omega} & c_{\omega}k_{xn} t_{ck} & -c_{\omega}k_{zn} t_{ck} \\
c_{\omega}k_{xn} t_{ck} & s_{\omega} & 0 \\
-c_{\omega}k_{zn} t_{ck} & 0 & s_{\omega}
\end{bmatrix} \,, \\[5pt]
\vec{Q} & = {\hat{T}}
\begin{bmatrix}
 i k_{xn}\chi_{4}^{'} & -i k_{zn}\chi_{4}^{'} & 0 \\
 i k_{xn}^2 \chi_{4} & -i k_{xn} k_{zn} \chi_{4} & i k_{zn} s_{\omega} \\
-i k_{xn} k_{zn} \chi_{4} & ik_{zn}^2 \chi_{4} & i k_{xn} s_{\omega}
\end{bmatrix} \,,
\end{align}
\end{subequations}
with $k_{xn} = k_x / k$, $k_{zn} = k_z / k$, $\beta_0 = v_0 / c$, $c_{\omega} = \cos(\omega \Delta t/2)$, $s_{\omega} = \sin(\omega \Delta t/2)$, and
\begin{align}
% k_{xn} = \frac{k_x}{k} \,, \quad k_{zn} = \frac{k_z}{k} \,, \quad \beta_0 = \frac{v_0}{c} \,, \\
% c_{\omega} = \sin(\omega \Delta t/2) \,, \quad s_{\omega} = \cos(\omega \Delta t/2) \,, \\[5pt]
\ r_{\omega \nu} = \frac{\theta^*}{\sqrt{e_{\omega}}} \frac{A_2 (\theta^2-e^{i\omega\Delta t}) + \theta^2 (A_1 - \nu A_{\nu} )(e^{i\omega\Delta t}-1)}{ (1-\theta^2)} \,.
\end{align}

Even though the matrix $\boldsymbol{M_{av}}$ has multiple zeros entries, it is difficult to find an analytical solution of equation~(\ref{eq:supp-det}) for any pair $(k_x,k_z)$.
To solve it numerically, we used the secant method as a root-finding algorithm, which allowed us to calculate the NCI growth rates across a wide range of frequencies.

We remark that in the case of the standard Galilean PSATD scheme, $\boldsymbol{M_{av}}$ reduces to
\begin{equation}
\boldsymbol{M_{av}}=
\begin{bmatrix}
\boldsymbol{R} & \boldsymbol{Q} \\
\boldsymbol{P} & \boldsymbol{I} \\
\end{bmatrix} \,,
\end{equation}
which is equivalent to equation~(19) of \cite{LehePRE2016}.

\end{document}